\documentclass[english]{llncs}
\usepackage[T1]{fontenc}
\usepackage[latin9]{inputenc}
\usepackage{babel}
\usepackage{float}
\usepackage{url}
\usepackage{amsmath}
\usepackage{amssymb}
\usepackage[unicode=true,pdfusetitle,
 bookmarks=true,bookmarksnumbered=false,bookmarksopen=false,
 breaklinks=false,pdfborder={0 0 0},pdfborderstyle={},backref=false,colorlinks=false]
 {hyperref}

\makeatletter

\floatstyle{ruled}
\newfloat{algorithm}{tbp}{loa}
\providecommand{\algorithmname}{Algorithm}
\floatname{algorithm}{\protect\algorithmname}

\usepackage{bussproofs}
\floatname{algorithm}{Listing}
\def\translate{\ $\leadsto$\ }
\def\eval{\ \Downarrow\ }
\usepackage{subfig}
\AtBeginDocument{\newsubfloat{algorithm}}

\makeatother

\usepackage{listings}
\begin{document}

\title{Transpiling Programming Computable Functions to Answer Set Programs}

\author{Ingmar Dasseville, Marc Denecker}

\institute{KU Leuven, Dept. of Computer Science, B-3001 Leuven, Belgium.\\
\texttt{ingmar.dasseville@cs.kuleuven.be}~\\
\texttt{marc.denecker@cs.kuleuven.be}}
\maketitle
\begin{abstract}
Programming Computable Functions (PCF) is a simplified programming
language which provides the theoretical basis of modern functional
programming languages. Answer set programming (ASP) is a programming
paradigm focused on solving search problems. In this paper we provide
a translation from PCF to ASP. Using this translation it becomes possible
to specify search problems using PCF.
\end{abstract}

\section{Introduction}

A lot of research aims to put more abstraction layers into modelling
languages for search problems. One common approach is to add templates
or macros to a language to enable the reuse of a concept \cite{iclp/DassevilleHBJD16,tplp/DassevilleHJD15,nmr/IanniIPSC04}.
Some languages such as HiLog \cite{jlp/Chen93} introduce higher order
language constructs with first order semantics to mimic this kind
of features. While the lambda calculus is generally not considered
a regular modelling language, one of its strengths is the ability
to easily define abstractions. We aim to shrink this gap by showing
how the lambda calculus can be translated into existing paradigms.
Our end goal is to leverage existing search technology for logic programming
languages to serve as a search engine for problems specified in functional
languages.

In this paper we introduce a transpiling algorithm between Programming
Computable Functions, a programming language based on the lambda calculus
and Answer Set Programming, a logic-based modelling language. This
transpilation is a source-to-source translation of programs. We show
how this can be the basis for a functional modelling language which
combines the advantages of easy abstractions and reasoning over non-defined
symbols. Transpiling Programmable Computable Functions to other languages
like C \cite{PCF2C}, or a theorem prover like Coq \cite{DBLP:conf/lpar/DargayeL07}
has been done before. But as far as the authors are aware, no approaches
to translate it into logic programs have been done before.

In Section \ref{sec:Programming-Computable-Functions} we introduce
the source language of our transpiler: Programming Computable Functions.
In Section \ref{sec:Answer-Set-Programming} we introduce the target
language of our transpiler: Answer Set Programming. In Section \ref{sec:Translation}
we describe the translation algorithm. Finally, in Section \ref{sec:Applications}
we motivate why this kind of translation can be useful in practice.
An implementation of the translator is made available online (\url{https://dtai.cs.kuleuven.be/krr/pcf2asp}),
where it can be tested with your own examples.

\section{Programming Computable Functions\label{sec:Programming-Computable-Functions}}

Programming Computable Functions \cite{DBLP:series/utcs/DowekL11,DBLP:journals/tcs/Plotkin77}
(PCF) is a programming language based on the lambda calculus. It is
not used as an end-user language; instead it provides a strong theoretical
basis for more elaborate languages, such as Lisp, Caml or Haskell.
There are many small variations of PCF, some extend it with booleans,
tuples or arithmetic operators. One such variation is known as Mini-ML
\cite{DBLP:conf/lfp/ClementDDK86}. The particular flavor is irrelevant
for the principles in this paper.

\subsection{Syntax}

The syntax of PCF relies heavily on the standard lambda calculus,
extended with natural numbers, a selection construct and a fixpoint
operator. We identify the following language constructs:
\begin{itemize}
\item function application \lstinline!$e_1 e_2$!, which is left associative,
\item a lambda abstraction \lstinline!$\lambda$x.$e$!, abstracting the
variable \lstinline!x! out of the expression \lstinline!e!,
\item for each numeral $n\in\mathbb{N}$, a constant \lstinline!n!,
\item constants \lstinline!succ!, representing the successor function over
$\mathbb{N}$, \lstinline!pred! representing the predecessor function
over $\mathbb{N}$,
\item a constant \lstinline!fix!, representing the fixpoint operator, also
known as the Y-combinator, and
\item a ternary language construct \lstinline!ifz $e_z$ then $e_t$ else $e_e$!,
representing an if zero-then-else.
\end{itemize}
Suppose that $\mathbb{\mathbb{I}}$ is an infinite supply of identifiers.
The syntax of PCF can be inductively defined as:

\begin{lstlisting}[language=Prolog,basicstyle={\ttfamily},breaklines=true,showstringspaces=false]
e = x ($\in \mathbb{I}$)  | e e  | $\lambda$x.e 
  | $n$ ($\in \mathbb{N}$) | succ | pred | fix | ifz e then e else e
\end{lstlisting}

\begin{example}
\lstinline!($\lambda$x. succ (succ x)) (succ 0)! is a complicated
way to write $3$. 
\end{example}

\noindent The expression \lstinline!fix! allows us to write functions
which would require recursive definitions in most programming languages.
It takes a function $f$ as argument and returns the fixpoint x of
that function so that $f(x)=x$. From this it follows that \lstinline!fix!
satisfies the equation \lstinline!fix $f$  = $f$ (fix $f$)!.
\begin{example}
A traditional recursive definition for the double of a number \lstinline!x!
could be: 

\lstinline!double x = ifz x then 0 else 1 + 1 + double (x-1)!

\noindent It is possible to rewrite this using fix, by abstracting
both \lstinline!double! and \lstinline!x!, and using \lstinline!pred!
and \lstinline!succ! for the increments and decrements: 

\lstinline!fix ($\lambda$double. $\lambda$x. ifz x then 0 else succ (succ (double (pred x))!

\noindent The informal meaning of this expression is the doubling
function.
\end{example}

\begin{example}
\lstinline!fix ($\lambda$plus. $\lambda$a. $\lambda$b. ifz a then b else plus (pred a) (succ b))!
of which the informal meaning is the binary sum function over natural
numbers.
\end{example}

\subsection{Operational Semantics}

When considering expressions, we traditionally consider only those
without free variables. However, when considering the operational
semantics, we will generalise this to situations where free variables
can occur. For this reason we introduce environments and closures
through a mutually inductive definition.
\begin{definition}
An \emph{environment} E is a mapping from identifiers to closures.
A \emph{closure} (E,e) consists of an environment E and an expression
e, where the environment must interpret at least all the free variables
in e.
\end{definition}

\noindent We say an environment interprets an identifier x if it
contains a mapping for x. The closure to which E maps an interpreted
variable x is written as E{[}x{]}.
\begin{example}
(\lstinline!succ a!,\{\lstinline!a! $\mapsto$(\{\},\lstinline!succ 0!)\})
is a valid closure which will evaluate to the number 2. 
\end{example}

\subsubsection{Evaluation context}

The evaluation relation $\eval$ is a relation between closures and
values, which we will write as follows:

\begin{prooftree}
\AxiomC{E, $e$ $\eval$ V}
\end{prooftree}
\begin{description}
\item [{(E,e)}] is the closure that is being evaluated. When considering
the evaluation of an expression without an explicit environment, we
assume it has no free variables and we interpret this is as the closure
with the empty environment.
\item [{V}] is the value that corresponds to the expression, this can either
be a natural number or a closure. A natural number can be implicitly
used as a closure with the empty environment.
\end{description}

\subsubsection{Notation}

We will describe both the semantics of PCF and the translation algorithm
using a set of inference rules. These are rules of the form

\begin{prooftree}
\AxiomC{Premise$_1$}
\AxiomC{...}
\AxiomC{Premise$_n$}

\TrinaryInfC{Conclusion}

\end{prooftree}

\noindent An algorithmic interpretation of these rules will lead
to a program which can evaluate/translate PCF. Most often, the easiest
way to read this kind of rules is bottom up.

\subsubsection{Evaluation Rules}

The following inference rules determine the operational semantics
for PCF through the evaluation relation $\eval$.

\begin{prooftree}

\AxiomC{E{[}\lstinline!x!{]}=(E$_{2}$,$e$)}

\AxiomC{ E$_{2}$, $e$$\eval$ $V$}

\BinaryInfC{E, \lstinline!x! $\eval$ $V$}

\end{prooftree}

\begin{prooftree}

\AxiomC{E, $e_{1}$ $\eval$(E$_{2}$, \lstinline!$\lambda$x.$e_3$!)}

\AxiomC{E, $e_{2}$ $\eval$$V$}

\AxiomC{$\text{E}_{2}\cup\{$\lstinline!x! $\mapsto V\}$, \lstinline!$e_3$!
$\eval$$V_{ap}$}

\TrinaryInfC{E, $e_{1}e_{2}$ $\eval$$V_{ap}$}

\end{prooftree}

\begin{prooftree}
\AxiomC{}

\UnaryInfC{E, \lstinline!$\lambda$x.$f$! $\eval$(E, \lstinline!$\lambda$x.$f$!)}

\end{prooftree}

\begin{prooftree}
\AxiomC{}
\UnaryInfC{E, $n (\in \mathbb{N}) \eval n$}

\end{prooftree}

\begin{prooftree}

\AxiomC{E, $e$ $\eval$$n$}

\UnaryInfC{$E$, \lstinline!succ $\ e$! $\eval$$n+1$}

\end{prooftree}

\begin{prooftree}

\AxiomC{E, $e$ $\eval$$n+1$}

\UnaryInfC{E, \lstinline!pred $\ e$! $\eval$$n$}

\end{prooftree}

\begin{prooftree}

\AxiomC{E, $e_{i}$$\eval$$0$}

\AxiomC{E, $e_{t}$$\eval$$V$}

\BinaryInfC{E, \lstinline!ifz $\ e_z\ $ then $\ e_t\ $ else $\ e_e\ $!$\eval V$}

\end{prooftree}

\begin{prooftree}

\AxiomC{E, $e_{i}$$\eval$$n$}

\AxiomC{$n>0$}

\AxiomC{E, $e_{e}$$\eval$$V$}

\TrinaryInfC{E, \lstinline!ifz $\ e_z\ $ then $\ e_t\ $ else $\ e_e\ $!$\eval V$}

\end{prooftree}

\begin{prooftree}

\AxiomC{$\text{E}\cup\{$\lstinline!x!$\mapsto$(E, \lstinline!fix ($\lambda$x. $e$)!)$\}$,
\lstinline!$e$! $\eval$$V$}

\UnaryInfC{E, \lstinline!fix ($\lambda$x. $e$)! $\eval$$V$}

\end{prooftree}

\noindent These rules form an inductive definition of the evaluation
relation $\eval$. Note that this is a call-by-value semantics. This
can be seen in the rule of applications, as the subexpression $e_{2}$
is evaluated before adding it to the environment. A call-by-name semantics
would just add the closure containing $e_{2}$ instead of the evaluation
of $e_{2}$.
\begin{example}
In the below tree you can follow the semantics of an expression using
multiple inference rules. Every horizontal line represents the application
of one evaluation rule.
\end{example}

\begin{prooftree}
\AxiomC{}

\UnaryInfC{\{(\lstinline!f! $\mapsto$\lstinline!($\emptyset$,fix ($\lambda$f. 4))!\},
\lstinline!4!$\eval4$}

\UnaryInfC{$\emptyset$, \lstinline!(fix ($\lambda$f. 4))!$\eval4$}

\AxiomC{$4>0$}

\AxiomC{}

\UnaryInfC{$\emptyset$,\lstinline!2!$\eval$2}

\AxiomC{}

\UnaryInfC{$\{$\lstinline!x!$\mapsto$2$\}$,\lstinline!2!$\eval$2}

\UnaryInfC{$\{$\lstinline!x!$\mapsto$2$\}$,\lstinline!x!$\eval$2}

\UnaryInfC{$\{$\lstinline!x!$\mapsto$2$\}$,\lstinline!pred x!$\eval$1}

\BinaryInfC{$\emptyset$,\lstinline!($\lambda$x. pred x) 2!$\eval$1}

\TrinaryInfC{$\emptyset$, \lstinline!ifz (fix ($\lambda$f. 4)) then 3 else ($\lambda$x. pred x) 2!$\eval1$}

\end{prooftree}

\section{Answer Set Programming\label{sec:Answer-Set-Programming}}

Answer Set Programming \cite{DBLP:journals/cacm/BrewkaET11} (ASP)
is a modelling language with a strong basis in logic programming.
It is mainly used as a language to specify NP-hard search problems
\cite{jelia/Niemela06}. There are a lot of different systems supporting
a unified ASP standard\cite{aspcore2}. An ASP program is essentially
a logic program with some extra syntactic restrictions. An ASP solver
computes the answer sets of the program under the stable semantics.
An answer set consists of a set of atoms which together represent
the solution of a problem. One program may have zero, one or multiple
answer sets.

\subsection{Language}

An ASP program is a set of rules of the form:
\begin{lstlisting}
head :- body$_1$,$\ldots$,body$_n$,not body$_{n+1}$,$\ldots$,not body$_{m}$
\end{lstlisting}

\noindent The first $n$ body atoms are positive, the others are
negative. The head and body atoms of the rules are of the form \lstinline!id(term$_1,\ldots,$term$_n$)!.
Body atoms can also be comparisons (<,>,=,$\neq$) between terms.
Terms can be either constants, variables, or arithmetic expressions
over terms. Constants are numbers or named constants (strings starting
with a lowercase character). Variables are represented as strings
starting with an uppercase character. An ASP program is considered
\emph{safe} if all rules are safe. A rule is considered safe if all
variables occurring in the rule, occur at least once in a positive
body. If the head is omitted, the rule is considered a constraint.
In this case no instantations of the body of the rule should exist
such that all the bodies are true. 

Choice rules are a common syntactic extension for ASP. These allow
heads of the form \lstinline!c$_l$ {a(X) : b(X) } c$_u$!, where $c_{l},c_{u}\in\mathbb{N}$
and $c_{l}\leq c_{u}$. This head is considered true if between $c_{l}$
and $c_{u}$ instances of \lstinline!a(X)! are true, given \lstinline!b(X)!.
They allow to easily introduce symbols that are not uniquely defined.
We can for instance declare p to be a singleton containing a number
between 1 and 10 with the choice rule: \lstinline!1 {p(X) : X = 1..10 } 1!.
The ASP program containing only this line has 10 answer sets, one
for each possible singleton.

\newsavebox{\firstlisting}
\begin{lrbox}{\firstlisting}
\begin{lstlisting}[numbers=left]
p(1). p(2). p(3). p(4).
1 {q(X) : p(X) } 2.
r(X + Y) :- q(X), q(Y).
:- not r(5).
\end{lstlisting}
\end{lrbox}
\newsavebox{\secondlisting}
\begin{lrbox}{\secondlisting}
\begin{lstlisting}
Answer Set 1:
p(1) p(2) p(3) p(4) 
q(1) q(4) r(2) r(5) r(8) 
Answer Set 2:
p(1) p(2) p(3) p(4) 
q(2) q(3) r(4) r(5) r(6) 
\end{lstlisting}
\end{lrbox}

\begin{algorithm}
  \subfloat[An ASP Program]{\usebox{\firstlisting}} \hfill%
  \subfloat[The Answer Sets]{\usebox{\secondlisting}}

\caption{\label{alg:ASP-Example}An example ASP program and its solutions}
\end{algorithm}

\begin{example}
In Listing \ref{alg:ASP-Example} you can see an example ASP program
together with its answer sets. The first line of the program defines
the predicate p as the numbers between 1 and 4. The second line is
a choice rule with no bodies. It states that q is a subset of p and
contains 1 or 2 elements. The third line says that r is the sum of
any two elements (possibly the same one) from q. The fourth line asserts
that r should contain 5. 
\end{example}

\subsection{Grounding (and Solving)\label{subsec:Grounding-(and-Solving)}}

To understand the details of the translation mechanism, basic knowledge
of how an ASP system constructs an answer set is needed. Constructing
answer sets happens in two phases: grounding and solving \cite{DBLP:journals/aim/KaufmannLPS16}.
The grounding process transforms the ASP program to an equivalent
propositional program. The solver then constructs the actual answer
sets from this propositional format. The translation from PCF described
in this paper will produce a fully positive, monotone theory without
choice rules or constraints. ASP grounders produce the actual (unique)
answer set for this kind of programs. Note that not all ASP systems
use the same algorithms, but the information presented here is common
to most systems.

The grounding process uses a bottom-up induction of the program. At
any point in time, the grounder contains a set of atoms which are
possibly part of an answer set. This set starts empty, and by the
end of the process this set contains an overapproximation of all answer
sets. The grounder tries to instantiate rules using this set of atoms.
Whenever a rule is instantiated, the instantiated head is added to
this set, and the ground instantiation of the rule is added to the
grounding of the program. ASP grounders require that all variables
occur in a positive body atom, this is the so-called safety requirement\emph{
}on rules. Safe rules have the property that only the positive part
of the program is essential for finding all rule instantiations and
current grounding approaches heavily rely on this property.
\begin{example}
Consider the rule \lstinline!d(X-1) :- d(X), X > 0! and the current
set of grounded atoms is just the singleton \{\lstinline!d(1)!\}.
The grounder can now instantiate the body atom \lstinline!d(X)! with
\lstinline!X!$=1$. The other body atom $1>0$ can be statically
evaluated to be true. This leads to the newly ground rule \lstinline!d(0) :- d(1)!
and \lstinline!d(0)! is added to the set of grounded atoms. The grounder
can now try to instantiate the rule with \lstinline!X!$=0$, but
the comparison $0>0$ prevents the rule to be added to the ground
program.
\end{example}

\noindent After the grounding phase an ASP solver can produce the
actual answer sets based on the grounding. An ASP solver typically
uses a SAT solver extended with some ASP specific propagators. The
inner workings of these programs are not needed to understand the
contents of this paper.

\section{Translation\label{sec:Translation}}

In this section we explain the core of the translation mechanism.
Section \ref{subsec:Characterisation-of-the} defines the relation
between the translation and the PCF semantics. Section \ref{subsec:Conventions}
introduces some conventions which explain the structure of the resulting
program. Finally, Section \ref{subsec:Static-Preamble} explains the
static part of the translation. Section \ref{subsec:Translation-Algorithm}
defines the translation relation between PCF expressions and the dynamic
part of the translation.

\subsection{Characterisation of the translation\label{subsec:Characterisation-of-the}}

\subsubsection{Translation relation}

The translation is characterised using a relation \translate which
we will write as follows:

\begin{prooftree}
\AxiomC{(E,$S_1$), $e$ \translate $A$, (t,$S_2$)}
\end{prooftree}
\begin{description}
\item [{E}] is a mapping from PCF-variables to ASP-terms for at least the
free variables in $e$. This works analogously to the environment
of the PCF semantics, which was the mapping from PCF-variables to
closures.
\item [{S$_{1}$}] is a set of ASP atoms ensuring the ASP-terms in E are
safe, and constraints enforcing the ifzero-semantics. 
\item [{e}] is the PCF expression that is translated. 
\item [{A}] is the ASP program consisting of a set of safe ASP rules, this
is the program that contains all the helper rules to translate $e$
\item [{t}] is the ASP term which represents the translation of $e$
\item [{S$_{2}$}] is the set of ASP atoms ensuring that $t$ is safe
\end{description}
It can be unintuitive that there are ASP-terms occuring on both sides
of the translation relation. The explanation for this lies in the
handling of free variables. The translation relation will be defined
structurally, this means that for the translation of a composite term,
the translation of its subterms are needed. This implies that when
translating the expression \texttt{}\lstinline!($\lambda$x. x))!\texttt{,
}the subterm \texttt{}\lstinline!x!\texttt{ }needs to be translated
as well. The translation needs some context to interpret this \texttt{}\lstinline!x!
and the context of a translation environment will be some information
about the parts which are already translated. 

A PCF expression corresponds to a single value, but a logic program
corresponds to an answer set with a lot of atoms. We need a way to
indicate the actual value that is meant with the logic program. The
$result$-predicate is used to indicate the resulting value of the
program.
\begin{definition}
The ASP translation of a PCF expression $e$ determined by \translate
is the ASP program $A$ such that $(\emptyset,\emptyset),e\text{\translate}A_{2},(t,S)$
and $A=A_{2}\cup$\texttt{\{}\lstinline!result(t):-S!\}.
\end{definition}

\subsubsection{Soundness of the translation}

PCF inherently works on expressions which evaluate to a particular
value, ASP programs define relations. A certain equivalence criterion
is needed to validate the translation. For this we use the $result$-predicate.
For ease of defining the correspondence the soundness criterion is
restricted to programs with a numeric evaluation.
\begin{definition}
A \emph{sound translator} for PCF to ASP maps every PCF expression
$e$ to an ASP program $A$ with a unique answer set. This answer
set contains at most one atom for the $result$-predicate. If $\emptyset,e\eval n\in\mathbb{N}$,
then $result(n)$ must be an element of the answer set of $A$.
\end{definition}

\begin{claim}
The translation of PCF expressions determined by \translate is a
sound translator.
\end{claim}

\noindent In this paper we will not prove this claim. We state it
here to give the reader an intuition about the correspondence between
a program and its translation.

\subsection{Conventions\label{subsec:Conventions}}

In the translation, all PCF expressions $e$ correspond to a tuple
$(t,S)$ where $t$ is an ASP term and $S$ is a set of ASP bodies.
Natural numbers have constants in both PCF and ASP which have a natural
correspondence. PCF functions are identified by an ASP term $t_{f}$,
so that for every ASP term $t_{x}$, the tuple $(Y,\{\mathit{inter}(t_{f},t_{x}),Y)\})$
denotes the image of the function $t_{f}$ applied to $t_{x}$. All
functions have infinite domains, and thus the full function cannot
be represented in a finite answer set. The $\mathit{domain}$ predicate
serves the purpose of making a finite estimate of the relevant parts
of the function. If at some point in the evaluation of $e$, the function
$t_{f}$ is applied to the value $t_{x}$, $\mathit{domain(t_{f},t_{x})}$
should be true. The $inter$-predicate only needs to be defined for
the domain of the function, resulting in a finite answer set containing
the relevant parts of the interpretation of the function.

Remember that $result$ predicate is used as the predicate determining
the final result of the program. So the translation of a PCF expression
is an ASP program, defining only 3 predicates: 
\begin{enumerate}
\item $\mathit{inter}$: determines the interpretation of functions
\item $\mathit{domain}$: determines the (relevant) domain of functions
\item $\mathit{result}$: determines the end result
\end{enumerate}

\subsubsection{Magic Set Transformation}

In these conventions a link with the magic set transformations \cite{DBLP:conf/sigmod/MumickFPR90}
of logic programs appears. The magic set transformation allows us
to transform a query, which is traditionally executed top-down, to
a program, which can be executed bottom-up. It uses the $magic$ predicates
to indicate which subqueries needs to be performed. As explained in
Section \ref{subsec:Grounding-(and-Solving)}, ASP uses a bottom-up
grounding process. So, the translation from PCF to ASP also converts
a top-down query (the evaluation of PCF) to a bottom-up process (the
ASP grounding). The $domain$-predicate has a function similar to
the $magic$ predicates: it indicates for which arguments a function
needs to be calculated.

\subsection{Static Preamble\label{subsec:Static-Preamble}}

The translation of any PCF expression consists of a dynamic part and
a static part. The static part ensures that the interpetation of the
\texttt{}\lstinline!succ!, \texttt{}\lstinline!pred! and \texttt{}\lstinline!fix!
builtins is taken care of. The dynamic part is produced by the translation
algorithm and takes care of the actual PCF expression. The static
part is the same for every translation and can be seen in Listing
\ref{alg:Static-part-of}. The first two lines of the static part
ensure the right translation of the \texttt{}\lstinline!pred! and
\texttt{}\lstinline!succ! terms. E.g. the PCF-term \texttt{}\lstinline!succ!
correctly corresponds to the ASP tuple $(\textit{succ},\{\})$ according
to the conventions defined in Section \ref{subsec:Conventions}. For
instance, if somewhere the PCF-term \texttt{}\lstinline!succ 0!
needs to be evaluated. The term will translate to $(Y,\{inter((succ,0),Y)\})$
which will result in $Y$ being equal to $1$ in the answer set.

Just like \texttt{}\lstinline!pred! and \texttt{}\lstinline!succ!,
the PCF- and ASP-term of \texttt{}\lstinline!fix! are the same.
But the required rules in the preamble are more complex. A naive translation
could look like this:

\begin{lstlisting}[language=Prolog,basicstyle={\ttfamily},breaklines=true,showstringspaces=false]
inter((fix,F),Z) :- inter((fix,F),Y), inter((F,Y),Z).
\end{lstlisting}
This rule most closely represents \lstinline!fix $f$  = $f$ (fix $f$)!,
but in the stable semantics this equation is not correctly represented
by the above rule. Instead, an intermediate term \texttt{}\lstinline!f(F)!
is introduced to symbolically represent the fixpoint of \lstinline!F!
in ASP. Now we are able to write the fixpoint as the function \lstinline!F!
applied to the symbolic function \lstinline!f(F)! as can be seen
on line 3. If the fixpoint is a function, we need to be able to apply
it to arguments. Line 4 serves this purpose: to apply \lstinline!X!
to a fixpoint of a function, you can apply \lstinline!F! to this
fixpoint (to ensure we do not have the symbolic representation) and
then apply \lstinline!X! to the result. Finally, lines 6 and 7 ensure
that the function applications performed in lines 3 and 4 are all
well-defined through the domain predicates. 

\begin{algorithm}
\begin{lstlisting}[language=Prolog,numbers=left,basicstyle={\ttfamily},breaklines=true,showstringspaces=false]
inter((pred,X),X-1) :- domain(pred,X), X > 0.
inter((succ,X),X+1) :- domain(succ,X).
inter((fix,F),Y)    :- domain(fix,F), inter((F,f(F)),Y).
inter((f(F),X),Y)   :- domain(f(F),X), inter((F,f(F)),FIX), 
                        inter((FIX,X),Y).
domain(F,f(F))      :- domain(fix,F).
domain(FIX,X)       :- domain(f(F),X), inter((F,f(F)),FIX).
\end{lstlisting}

\caption{\label{alg:Static-part-of}Static preamble of the ASP translation}
\end{algorithm}

\subsection{Translation Algorithm\label{subsec:Translation-Algorithm}}

In this section we present the translation algorithm as a definition
for the translation relation \translate using inference rules. Sometimes
new ASP constants or variables are needed in the translation. We suppose
there is some global supply of those. We use the notation $head\leftarrow B$
for the ASP rule where $head$ is the head atom and $B$ is the set
of body atoms.

\subsubsection{Scoping}

When translating an expression, the free variables in this expression
need to be filled in. As we translate nested expressions level per
level, we need to pass these values along the expression tree. For
this reason, we do not just associate an identifier with a function
but a tuple containing an identifier and the current scope. The current
scope is a tupling of the full codomain of the translation environment
$E$. We will refer to it as $\mathit{scope}_{E}$ .

\subsubsection{Builtins (numbers, pred, succ, fix)}

\begin{prooftree}
\AxiomC{}
\UnaryInfC{(E,S), $b$ \translate $\emptyset$, ($b$,S)}
\end{prooftree}Builtins are relatively easy to translate. The hard work is taken
care of by the static preamble described in Section \ref{subsec:Static-Preamble}.
A builtin produces no new ASP rules and is translated by itself. Safety
is however taken into account, not for the scoping of variables, but
for the handling of the if-zero constraints.

\subsubsection{Variable}

\begin{prooftree}

\AxiomC{}

\UnaryInfC{(E,S), \lstinline!x! \translate $\emptyset$, (E{[}\lstinline!x!{]},S)}

\end{prooftree}

The algorithm carries around a mapping that represents how variables
should be translated. This makes translating it a simple variable
easy: just look it up in the mapping and combine it with the required
safety.

\subsubsection{Application}

\begin{prooftree}

\AxiomC{(E,S), $e_{1}$ \translate $A_{1}$, $(t_{1},B_{1})$}

\AxiomC{(E,S), $e_{2}$ \translate $A_{2}$, $(t_{2},B_{2})$}

\BinaryInfC{(E,S), \lstinline!$e_1$$e_2$! \translate $A_{1}\cup A_{2}\cup rule_{domain}$,
$(X,\mathit{body}_{inter}\cup B_{1}\cup B_{2})$}

\end{prooftree}

\begin{alignat*}{1}
X= & \textit{a new ASP variable}\\
\mathit{rule}_{domain}= & \{\textit{domain}(t_{1},t_{2})\leftarrow B_{1}\cup B_{2}\}\\
\mathit{body}_{\mathit{inter}}= & \{inter((t_{1},t_{2}),X)\}
\end{alignat*}
Applications are translated by independently translating the two subexpressions.
The produced ASP programs need to be combined, with the additional
rule that $t_{2}$ should be added to the domain of the function $t_{1}$.
To obtain the resulting value, we use the $inter$-predicate according
to the conventions explained in Section \ref{subsec:Conventions}.
\begin{example}
The rule below shows how the application rule can be used to translate
the successor of 1. The static part of the translation ensures that
the \lstinline!inter!-relation for \lstinline!succ! is interpreted
correctly so that in any solution. The $X$ gets evaluated to 2.

\begin{prooftree}

\AxiomC{($\emptyset$,$\emptyset$), \lstinline!succ! \translate
$\emptyset$, $($\lstinline!succ!$,\emptyset)$}

\AxiomC{($\emptyset$,$\emptyset$), 1 \translate $\emptyset$,
$(1,\emptyset)$}

\BinaryInfC{($\emptyset$,$\emptyset$), \lstinline!succ 1! \translate
$\{domain(succ,1)\}$, $(X,inter((succ,1),X))$}

\end{prooftree}
\end{example}

\subsubsection{Lambda}

\begin{prooftree}
\AxiomC{(E $\cup ($\lstinline!x!$,X)$,$S\cup body_{domain})$, $e$ \translate $A$, $(t,B)$ }
\UnaryInfC{(E,S), \lstinline!$\lambda$x. e! \translate $A\cup rule_{inter}$ , $((l,scope_E), S)$}
\end{prooftree}

\begin{alignat*}{1}
X= & \textit{a new ASP variable}\\
l= & \textit{a new ASP constant}\\
\mathit{rule}_{inter}= & \{inter(((l,scope_{E}),X),t)\leftarrow B\}\\
\mathit{body}_{\mathit{domain}}= & \{\mathit{domain}((l,scope_{E}),X)\}
\end{alignat*}
Lambda expressions bring a new variable into scope, so they modify
the $(E,S)$-environment before recursively translating the body of
the expression. The freshly scoped variable needs to be put into the
scoping function $E$, for this we assign it a new ASP variable ($X$
in the rule). This variable should have a finite range, we invent
a new name for our function ($l$ in the rule) and use the $\mathit{domain}$
predicate to restrict $X$ to the domain of the function. The resulting
translation $(t,B)$ represents the image of the function, so the
rule $\mathit{rule}_{inter}$ is added to couple the representation
of the function to its interpretation.
\begin{example}
$(\emptyset,\emptyset)$\lstinline!($\lambda$x.2)!\translate\{$inter(((\mathit{l},()),X),2)\leftarrow\mathit{domain}((l,()),X)$\},$((l,()),\emptyset)$\\
This can be read as follows: The translation of the constant function
to 2 in an empty environment is represented by the constant $(l,())$.
The interpretation of $(l,())$ when applied to any term $X$ in the
domain of $(l,())$ is 2.
\end{example}

\subsubsection{If zero-then-else}

\begin{prooftree}

\AxiomC{(E,S), \lstinline!e$_{\texttt{ifz}}$! \translate $A_{\texttt{ifz}}$, $(t_{\texttt{ifz}},B_{\texttt{ifz}})$}
\noLine
\UnaryInfC{$(E, B_{\texttt{ifz}} \cup \{t_{\texttt{ifz}} = 0\})$, \lstinline!e$_{\texttt{then}}$! \translate $A_{\texttt{then}}$, $(t_{\texttt{then}},B_{\texttt{then}})$}
\noLine
\UnaryInfC{$(E, B_{\texttt{\texttt{ifz}}} \cup \{t_{\texttt{ifz}} \neq 0\})$, \lstinline!e$_{\texttt{else}}$! \translate $A_{\texttt{else}}$, $(t_{\texttt{else}},B_{\texttt{else}})$}
\UnaryInfC{(E,S), \lstinline!if e$_{\texttt{ifz}}\ $ then e$_{\texttt{then}}\ $ else e$_{\texttt{else}}$! \translate 
$A_{\texttt{ite}} \cup\ rule_{\texttt{ite}}, (X,S \cup body_{\texttt{ite}})$}
\end{prooftree}

\begin{alignat*}{1}
X= & \textit{a new ASP variable}\\
ite= & \textit{a new ASP constant}\\
\mathit{rule}_{ite}= & \{inter((\mathit{ite},scope_{E}),t_{\texttt{then}})\leftarrow B_{\texttt{then}}\\
 & ,\,inter((\mathit{ite},scope_{E}),t_{\texttt{else}})\leftarrow B_{\texttt{else}}\}\\
\mathit{body}_{\mathit{ite}}= & \{inter((\mathit{ite},scope_{E}),X)\}\\
A_{\texttt{ite}}= & \ensuremath{A_{\texttt{ifz}}\cup A_{\texttt{then}}\cup A_{\texttt{else}}}
\end{alignat*}
If zero expressions are translated using the translations of its three
subexpressions. But we need to alter the safety to ensure that the
``then''-part is only evaluated if the ``if''-part is 0 (and the
analog for the ``else'' part). To construct the value of the full
expression we define an intermediate symbol ($\mathit{ite}$ in the
rule) to represent the union of the ``then'' and the ``else''
part. Because the extra safety ($=0$, $\neq0$) is mutually exclusive,
only one of those terms will have a denotation, so the interpretation
of $\mathit{ite}$ will be unique.
\begin{example}
The translation of \texttt{}\lstinline[mathescape=true]!($\lambda$x. ifz x then succ else pred) 2 4!
is visible in Listing \ref{alg:Example-Translation}. The static part
is omitted. Lines 1 and 2 are result of the if zero-then-else translation.
Line 3 is the result of the lambda translation. Lines 4 and 5 are
the result of the application. And in line 6 the end result can be
seen. This rule can be read as follows: Let X2 be the application
of the function to 2. Let X3 be application of X2 to 4. The final
result is X3.
\end{example}

\begin{algorithm}
\begin{lstlisting}[language=Prolog,numbers=left,basicstyle={\ttfamily},breaklines=true,showstringspaces=false]
inter((ite1,(X0)),succ):-domain((l0,()),X0),X0=0.
inter((ite1,(X0)),pred):-domain((l0,()),X0),X0<>0.
inter(((l0,()),X0),X1):-domain((l0,()),X0),inter((ite1,(X0)),X1).
domain((l0,()),2).
domain(X2,4):-inter(((l0,()),2),X2).
result(X3):-inter(((l0,()),2),X2),inter((X2,4),X3).
% omitted static part visible in Listing 1
\end{lstlisting}

\caption{\label{alg:Example-Translation}Translation of \texttt{}\lstinline[mathescape=true]!($\lambda$x. ifz x then succ else pred) 2 4!}
\end{algorithm}

\subsection{Optimisations\label{subsec:Optimisations}}

The translation algorithm which is given in the previous section is
not an optimal translation. A lot of optimisations are possible, for
instance, not all variables in scope need to be present in $\mathit{scope}_{E}$,
only the ones which are actually used in the subexpression. Applying
such optimisations can significantly reduce the size of the grounding
of the ASP program. The possibilities here are very interesting research
topics, but are considered out of scope for this paper.

\subsection{Implementation}

An implementation was made in Kotlin. The runtime uses Clingo \cite{DBLP:journals/corr/GebserKKS14}
to run the resulting ASP files, but the resulting specifications could
be used with any ASP-Core-2 \cite{aspcore2} compliant system. On
\url{https://dtai.cs.kuleuven.be/krr/pcf2asp} you can find a tool
on which you can try out the translation. A few example PCF formulas
are provided, but you can ask for translations of arbitrary PCF formulas
and see their corresponding answer set. 

\section{Applications\label{sec:Applications}}

\subsection{Multiple interpretations for one variable}

Directly translating PCF gives us little more than a traditional interpreter
of PCF would do, but based on this translation we can provide extra
functionality, leveraging the existing ASP solvers. Traditional PCF
does not support the possibility that the interpretation of a term
is not uniquely defined, but we can extend PCF so we can declare the
variable $a$ as a number between 1 and 10 without defining its specific
value. In that case we can get (at most) 10 different evaluations
of our program, one for each interpretation of $a$. It is easy to
extend the translation to encode this in ASP. 

Traditional interpreters solve the question: ``What is the evaluation
of this program?''. But using these variables another question can
be interesting: what value(s) for $a$ should I choose so that the
program evaluates to 0. We can leverage the strengths of ASP solvers
to find the solutions. Expressing that the evaluation should be zero
can be done through a simple ASP constraint:

\begin{lstlisting}
:- not result(0).
\end{lstlisting}

When this constraint is added, the resulting answer sets will now
all have the same interpretation (0) for the $result$ predicate,
but we are interested in the interpretation for $a$.
\begin{example}
\label{exa:equation}In Listing \ref{alg:Uncertainty} you can see
a PCF expression representing that $a+b=c$. If we now use choice
rules in ASP to translate these variables to the domain of natural
numbers between 0 and 10, we can use ASP to find multiple solutions
of this equation. An example of how this would look in ASP an be seen
in Listing \ref{alg:choiceASP}.
\end{example}

\begin{algorithm}
\begin{lstlisting}[language=Prolog,numbers=left,basicstyle={\ttfamily},breaklines=true,showstringspaces=false]
($\lambda$eq. $\lambda$plus. 
      eq (plus a b) c)
    
(fix ($\lambda$eq.$\lambda$x.$\lambda$y. ifz x then (ifz y then 0 else 1) 
                        else (ifz y then 1 else eq (pred x) (pred y))))
(fix ($\lambda$plus.$\lambda$x.$\lambda$y. ifz y then x else plus (succ x) (pred y)))
\end{lstlisting}

\caption{\label{alg:Uncertainty}$a+b=c$ in PCF}
\end{algorithm}

\begin{algorithm}
\begin{lstlisting}[numbers=left,basicstyle={\ttfamily},breaklines=true,showstringspaces=false]
1{a(X)}1 :- X=1..10.
1{b(X)}1 :- X=1..10.
1{c(X)}1 :- X=1..10.
:- not result(0).
$\ldots$
domain(X1,A):-domain((l0,()),X0),domain((l1,(X0)),X1),a(A). 
$\ldots$
\end{lstlisting}

\caption{\label{alg:choiceASP}$a+b=c$ in ASP}
\end{algorithm}

\noindent  The problem in Example \ref{exa:equation} can easily
be generalised to arbitrarily complex polynomials to model mixed integer
problems. A graph coloring problem can be represented by using a new
constant for each node that needs to be colored and writing down an
expression that evaluated to 0 if the graph is colored correctly.

An important thing to note here is that ASP does not naively calculate
the result for all possible values of the choice rules. It uses a
CDCL-based solving algorithm to explore the search space in an intelligent
way. 

\subsection{Towards a more expressive language}

PCF is not intended to be an end-user language, but it serves as a
basis for many real world programming languages. Analogously, we are
developing a more expressive language based on the principles of PCF.
This language includes more complex data types for representations
which are more elegant than possible in PCF. Together with the multiple-model
semantics of ASP this leads to an interesting modelling language.
Using these ideas the new Functional Modelling System (FMS) is being
developed. On the website \url{https://dtai.cs.kuleuven.be/krr/fms}
a demonstration of this new system can be found. This system is an
extension of PCF with some more practical language constructs and
uses the translation principles described in this paper to use ASP
as a solver engine for this new language. However as indicated in
Section \ref{subsec:Optimisations}, a lot of optimisations are needed
to be competitive with native ASP encodings. The efficiency of these
translators have not been formally investigated yet.

\section{Conclusion}

We presented a translation from PCF to ASP programs. A basic translation
is easily implemented, and many optimisations are possible. With only
small changes, we can exploit the search power of ASP to solve problems
expressed in PCF. This translation can serve as a basis to use functional
programming techniques in modelling languages for search problems,
or even tighter integrations between functional and logical languages.
FMS is under development now and uses the techniques described in
this paper as a basis for its language.

\bibliographystyle{plain}
\bibliography{refs,krrlib}

\end{document}